\documentclass{article}
\usepackage{spconf,amsmath,graphicx}
\usepackage{booktabs, tabularx}
\newcolumntype{C}{>{\centering\arraybackslash}X}

\usepackage{mwe} 

\title{Ultrasound Confidence Maps of Intensity and Structure Based on Directed Acyclic Graph and Artifact Models}
 \name{Alex Ling Yu Hung$^{\dagger}$ \qquad Wanwen Chen$^{\star}$ \qquad John Galeotti$^{\dagger,\star}$}

\address{$^{\dagger}$ Department of Biomedical Engineering, Carnegie Mellon University, Pittsburgh PA, 15213, USA \\
     $^{\star}$ Robotics Institute, Carnegie Mellon University, Pittsburgh PA, 15213, USA}
\begin{document}
%
\maketitle
\begin{abstract}

Ultrasound imaging has been improving, but continues to suffer from inherent artifacts that are challenging to model, such as attenuation, shadowing, diffraction, speckle, etc.  These artifacts can
potentially confuse image analysis algorithms unless an attempt is made to assess
the certainty of individual pixel values. Our novel confidence algorithms analyze
pixel values using a directed acyclic graph based on acoustic physical properties of ultrasound imaging. We demonstrate unique capabilities of our approach and compare it against previous confidence-measurement algorithms for shadow-detection and image-compounding tasks.

\end{abstract}
\section{Introduction}
\label{sec:intro}
Ultrasound is a non-invasive, real-time and safe diagnostic imaging technique. However, it suffers from noise and artifacts, such as shadowing artifacts that depend on the direction of the probe and high attenuation coefficients of certain tissues. Ultrasound images are naturally brighter at the top and they tend to get darker as sound attenuates through deeper regions. Estimating the resulting uncertainty of individual pixel values can be helpful or essential for further image analysis.

Measuring uncertainty in ultrasound images has been discussed by many previous works, most of which were estimated the attenuation coefficients of the tissues in the images. For example, \cite{yu2010backscatter} compensated for artifacts and shadows and computed the map of attenuation coefficients by iteratively minimizing cost functions for back scatter, contour and attenuation. Other approaches utilized the image's corresponding raw Radio Frequency (RF) acoustic waveform data to estimate attenuation coefficients. Spectral analysis of RF data was used by \cite{treece2005ultrasound} to increase the locality and applicability of attenuation measurements.  To reduce system and transducer dependencies, \cite{yao1990backscatter} calculated the spectral difference by using a reference phantom to normalize the power spectra at different depth. Spectral cross-correlation was used by \cite{kim2007attenuation} to compare consecutive power spectra obtained from the backscattered RF signals at different depths to estimate the attenuation. A hybrid model \cite{kim2008hybrid} combined the strengths and mitigated the weaknesses of \cite{yao1990backscatter} and \cite{kim2007attenuation}. In earlier work, \cite{he1986attenuation} estimated the attenuation with the variance of the mean powers of the
overall echoes, and \cite{jang1988ultrasound} made use of the entropy difference between neighboring echo signals. Unlike other prior work, \cite{karamalis2012ultrasound} directly estimated the confidence of each pixel in ultrasound images without calculating the attenuation. This work made use of the random walk segmentation algorithm proposed by \cite{grady2006random}, and set the first row of the image as 1 and the last row as 0 in the graph, with the weight of each edge dependent on the image gradient. However, the algorithm deals with reverberation artifacts poorly and is sensitive to the change in intensity of speckle noise in images with fewer structures. 

Our confidence-map approach is inspired by \cite{karamalis2012ultrasound}, but we model the image differently to address key problems of previous approaches. Our main contributions are: (1) Modeling the confidence map in novel way which is more robust to abrupt changes in gradient in images with fewer structures, (2) A unique confidence measurement that better models diffraction and shadowing effects, (3) An innovative way to model speckle noise, needles and reverberation artifacts and (4) Proposal of a structural confidence that depicts the certainty of having a real anatomic structural boundary at the pixel.

\section{Methods}
\label{sec:methods}

\subsection{Speckle Noise Denoising}
Our confidence depends on the image gradient, but the speckle noise in the ultrasound images will make the gradient map noisy. It's important to remove such noise, so that we can model the attenuation better. Our speckle denoising algorithm is based on an anisotropic diffusion approach proposed by \cite{yu2002speckle}. They used an instantaneous coefficient of variation $q$ (Eq.~\ref{instantaneous_coefficient_of_variation}) to measure the homogeneity.
\begin{equation}
    \label{instantaneous_coefficient_of_variation}
    q=\sqrt{
    \frac{
        \frac{1}{2}(\frac{|\nabla I|}{I})^2 - \frac{1}{4^2}(\frac{\nabla^2I}{I})^2}
    {
    [1 + \frac{1}{4}(\frac{\nabla^2I}{I})]^2
    }
    }
\end{equation}
The diffusion coefficient $c(q)$ at every pixel is given by comparing the local $q$ and a reference $q_0$ in a known homogeneous region (Eq.~\ref{diffusion_coefficient}). Based on \cite{yu2002speckle}, we identify edges with large gradient with a Canny edge detector, and we then reduce those pixel's diffusion coefficient by a scaling factor $c_{canny}$,

\begin{equation}
    \label{diffusion_coefficient}
    c(q)=c_{canny} \frac{1}{1 + [q^2-q_0^2]/[q_0^2(1+q_0^2)]}
\end{equation}
After each iteration of diffusion, We match the histogram of the diffused image with the original image, to preserve 
the contrast and the brightness.
\subsection{Ultrasound Intensity Confidence}
\label{subsec:confidence}

\begin{figure}[htb]

  \centering
  \centerline{\includegraphics[width=8.5cm]{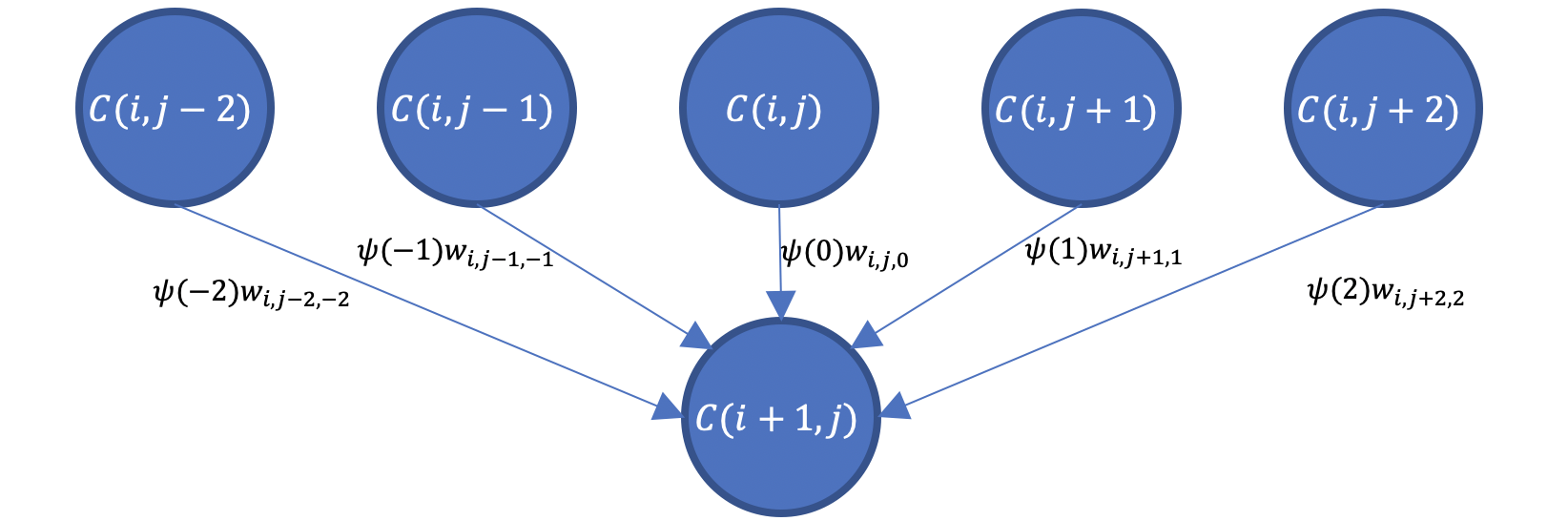}}
\caption{An example of the proposed model}
\label{model}
\end{figure}

Sound waves are emitted from the probe and propagate downwards. Along the way, the sound wave will be attenuated and distorted by the tissues. Our intensity confidence measurement depicts how sure we are in the pixel value based on the intensity of the sound waves, 
The confidence should fall off according to how much the intensity of the sound wave falls off. The intensity of the sound wave is path-dependent and is only related to intensity at the previous point and the attenuation between the previous point and the current point. Therefore, we model our confidence map as a directed graph, where the confidence of a row is only dependent on the confidence of the row above. To account for the diffraction effect of the sound wave, the confidence of each pixel not only depends on the confidence of the pixel that is right above it, but also depends on nearby pixels in the above row. This can also be viewed as a causal model where the confidence of the above row is the cause, and the current row is the effect. The noise random variables in the causal model are assumed to be given by the speckle noise which we removed earlier. Denote the confidence at pixel $(i,j)$ as $C(i,j)$, and the directed edge connecting pixel $(i,j)$ to pixel $(i+1,j')$ as $w_{i,j,j'-j}$, whose value is related to the image gradient and the depth of the pixel. An example of the proposed model is shown in Fig.~\ref{model}. 

We set the confidence value as $1$ in the first row of the image as the initialization, also can be interpreted as intervention in causal reasoning, then the confidence value at pixel $(i+1,j)$ is calculated from the following equation: 

\begin{equation}
C(i+1,j)=\sum_{k=-\kappa}^\kappa \psi(k)w_{i,j+k,k}C(i,j+k)
\end{equation}
where $\kappa$ indicates the range of pixels in the above row that can affect the current pixel. $\psi(k)$ is a weighting function that indicates how much effect the upper row has on the lower row with respect to the horizontal pixel distance $k$. In our case, the confidence directly above should contribute the most to the confidence of the pixel below, and further away preceding pixels should contribute less. We sample the weighting function $\psi(k)$ based on a normal distribution, 
\begin{equation}
\psi(k)= \begin{cases} 
      \Phi(\frac{k+0.5}{\sigma})-\Phi(\frac{k-0.5}{\sigma}) & k\neq \pm \kappa\\
      (1-\sum_{k=-\kappa+1}^{\kappa-1} \psi(k))/2& otherwise
   \end{cases}
\end{equation}
where $\Phi$ is the standard normal cumulative distribution.

The attenuation can be partially modeled by the gradient of the image, but in a naive approach noisy bright pixels at the top of the image would be inferred to cause attenuation. To alleviate the effects of noise, we use the relative gradient $g(i,j,d)$ as an alternative, where $i,j$ denotes the coordinates of the gradient and $d$ denotes the direction of the gradient. 
\begin{equation}g(i,j,d)=\frac{|I(i+1,j+d)-I(i,j)|}{\frac{1}{a-d}\sum_{k=0}^{a-d-1}|I(i+1,k+d)-I(i,k)|}\label{relative_gradient}\end{equation}
where $I$ is the ultrasound image and $a$ is the width of the image. The numerator of Eq.~\ref{relative_gradient} represents the gradient at $(i,j)$, and the denominator is the mean gradient of row $i$. 

Due to attenuation, the noise is less significant and the pixel values are much lower in the deeper region.  The proposed relative gradient might be undesirably large because the mean gradient is small. However, larger gradients deeper in the image will have less effect on the confidence values than shallow-region gradients. Inspired by \cite{karamalis2012ultrasound}, we use the Beer-Lambert Law \cite{swinehart1962beer} in a unique way. Denote $g'(i,j,d)$ as the Beer-Lambert-Law-adjusted relative gradient, 
\begin{equation}
g'(i,j,d)=g(i,j,d)^\beta e^{-\alpha\frac{i+1}{h}}
\end{equation}
where $h$ is the height of the image, $\alpha$ is attenuation coefficient, $\beta$ is the gradient adjusting factor.  

The gradient-dependent weight $w_{i,j,d}$ is then defined as
\begin{equation}
w_{i,j,d}=e^{-\gamma g'(i,j,d)}
\end{equation}
where 
\begin{equation}
\gamma=-\frac{ln\xi}{\sum_{i=1}^h e^{\alpha \frac{i}{h}}}
\end{equation}
The value of $\xi$ is set to be the desired confidence value in the hypothetical case of the bottom row of a completely homogeneous image; in this paper, $\xi=0.1$.

\subsection{Needle and Reverberation Artifacts Modeling}
Even though the proposed confidence measurement is capable of modeling the attenuation and diffraction effect, it doesn't take the reverberation by metallic objects such as needles 
into account. As the reverberation artifacts are artificial and mask underlying pixel values, those pixels should have very small confidence values. Metallic objects 
also attenuate the sound a lot more than other structures. Identifying needles (which may appear similar to anatomic structures) and reverberation artifacts can lead to better modeled confidence maps.

We utilize the needle and needle-reverberation artifact segmentation algorithm by \cite{hung}, to identify the needles and artifacts. We model the needle differently by modifying the relative gradient $g(i,j)$ for needle pixels, assigning the largest possible relative gradient to the edge on the needle and 1 to the rest of the needle. For pixel $(i,j)$ that belongs to a needle 
\begin{equation}
g(i,j)= \begin{cases} 
      \frac{g_m}{\frac{1}{a}\sum_{k=0}^{w-1}|I(i+1,k)-I(i,k)|} & (i,j)\in Edge\\
      1& otherwise
   \end{cases}
\end{equation}
where $g_m$ is the largest gradient value in the image. 

As for the reverberation artifact pixels, since they are purely artificial and don't interfere with the attenuation \cite{ziskin1982comet}, we simply set the relative gradient $g(i,j)=1$ for all artifact pixels $(i,j)$. When calculating the relative gradient, we also exclude the artifact pixels during the calculation of the mean, i.e. the denominator, since these artificial pixels are brighter. After the entire confidence map is calculated, we assign very low confidence value to the artifact pixels, because the reverberations are not caused by actual structures. Therefore, the final confidence map $\Tilde{C}(i,j)$ is given by 
$\Tilde{C}(i,j)=C(i,j)(1-Seg(i,j))$, 
where $Seg(i,j)$ is the output of the probabilistic artifact segmentation result.

\subsection{Ultrasound Structural Confidence}
The confidence map in Section~\ref{subsec:confidence} measures the confidence of each pixel value, but it does not assess the probability of having a real anatomic boundary at each pixel location. We propose a novel \emph{structural confidence map} that differentiates \emph{contrast} around real tissue boundaries vs. from artifacts. 

We begin by obtaining a reference intensity confidence map $R$ for our particular ultrasound system and its current settings, calculated based on ultrasound images of an empty phantom. Because there is no structure in the images to cause artifacts or occlusions, the confidence for each row in the reference map should be the maximum-possible confidence value for its distance from the transducer. Theoretically, when applying our approach to actual tissue, each value in the confidence map should be smaller than the value in the corresponding row in the reference map, since sound should be attenuated less in empty images. However, in practice, noise might change the behavior of the confidence map. To compensate for this, we set a constraint that while calculating an \emph{adjusted intensity confidence map}, $C'$, the confidence at a certain pixel could not be larger than the maximum confidence of the corresponding row in the reference map. We enforce this constraint by examining each value during the propagation of the confidence from top to bottom of the image, truncating confidence values that exceed reference-map values, and then continuing with confidence propagation to the row below. We denote the structural confidence map by $\Gamma(i,j)=C'(i,j)/R(i,j) \in [\,0.0, 1.0]\,$.  Pixels with lower ratios may be presumed to be under reflective surfaces where there are more likely to be artifacts and shadows. 

\section{Experiments and Discussion}
We evaluate our confidence-estimation methods on the tasks of (1) Identification of shadow and reverberation artifacts, (2) Detection of (partial) loss-of-contact between the (end of the) transducer and the skin surface and (3) Image compounding. 
The ultrasound imaging was performed with a UF-760AG Fukuda Denshi machine on chicken breast, a live pig and an anthropomorphic phantom 
produced by Advanced Medical Technologies. The code in this paper is run in python/NumPy on  Intel Core i5-8279U, where it takes 0.63s, 0.07s and 0.56s for \cite{karamalis2012ultrasound}, our intensity confidence without denoising, and our intensity confidence respectively to run on a $128\times128$ image, and 160.66s, 4.12s, and 40.99s on a $1024\times 1024$ image. 

\subsection{Shadow and Reverberation Artifact Detection}
\begin{figure}[htb]

  \centering
  \centerline{\includegraphics[width=8.5cm]{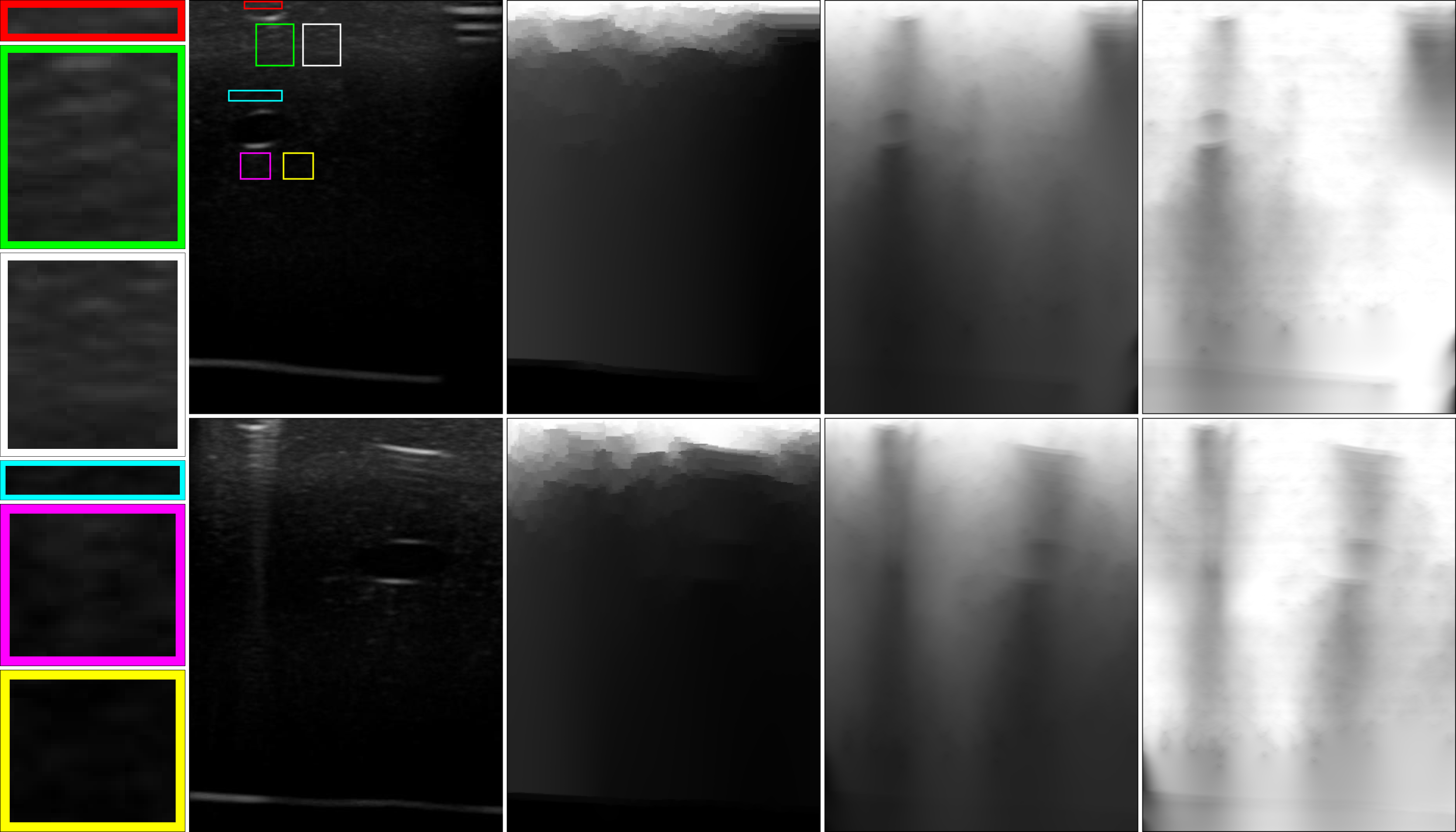}}
\caption{Left to right: zoom-in of the labeled patches, input image, confidence map by \cite{karamalis2012ultrasound}, our intensity confidence, our structural confidence. 
the red, green and white boxes are examples of $A$, $B$ and $C$ patch for reverberation artifacts. The blue, purple and yellow boxes are $A$, $B$ and $C$ for shadow. }
\label{comp1}
\end{figure}

\begin{figure}[htb]
  \centering
  \centerline{\includegraphics[width=8.5cm]{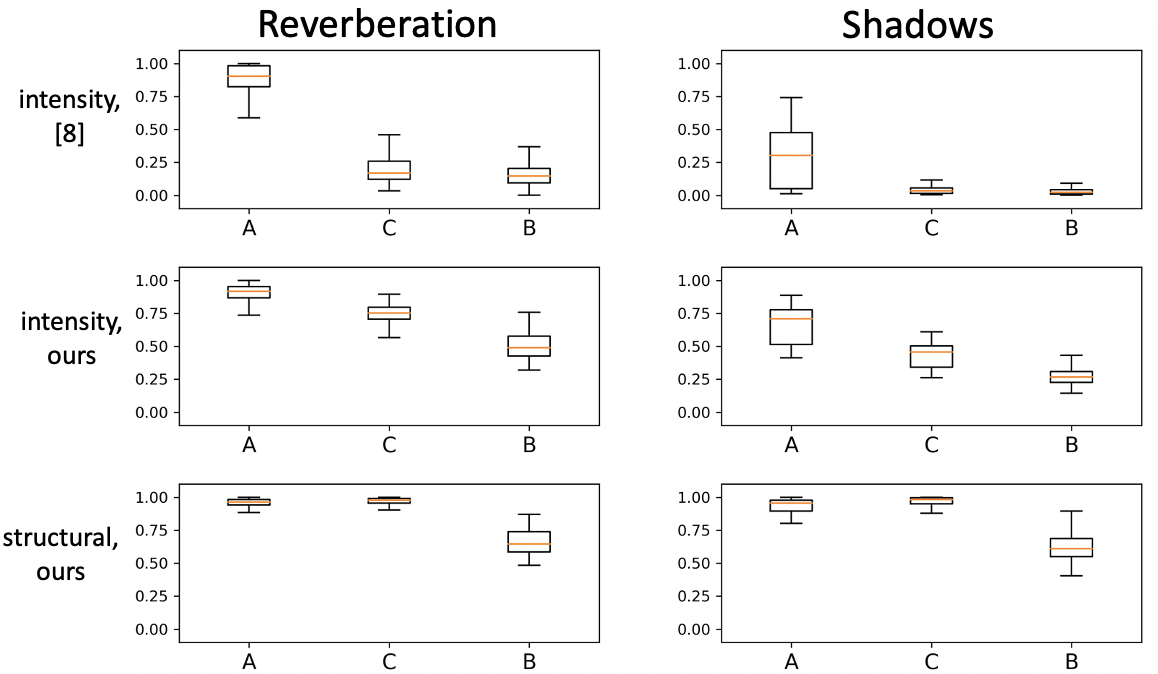}}
\caption{Comparison between \cite{karamalis2012ultrasound}, our intensity confidence, and our structural confidence in detecting reverberation artifacts and shadows, where the y-axis is the confidence value.
Our intensity confidence separates the different patches while \cite{karamalis2012ultrasound} can't, and our structural confidence successfully separates the artifact patch from non-artifact patch.}
\label{boxplot}
\end{figure}

In our initial experimental demonstration, we neither model needles nor reverberation artifacts nor shadows directly, testing our algorithm's generalized ability to infer appropriate confidence values for such challenging regions. We manually identified and labeled representative patches within 20 test images, to provide examples of image regions corresponding to artifacts, shadows, and adjacent regular tissues.  Referring to Fig. \ref{comp1}: patch $A$ is the region above actual tissue above the surface that causes the shadow or artifacts, patch $B$ is the shadowed region or artifact region and patch $C$ is another patch without shadows or artifacts in the same horizontal line with patch $B$. Visual (qualitative) results and comparisons are shown in Fig.~\ref{comp1}, where the overlays in the top left image shows how we labeled. The confidence map by \cite{karamalis2012ultrasound} is overly sensitive to abrupt changes in gradient within otherwise (semi)homogeneous image regions, leading to unnecessarily low confidence in much of the image. Our algorithm is more robust to such gradient change and produces intensity confidence values that decay slowly with depth when not in shadows or under artifacts.  Our method model the attenuation of sound well when it passes through the needle and vessel walls, giving the artifact and shadow lower value, as well as identifying the region where the probe is detached from the surface. The quantitative results are shown in Table.~\ref{tab} and Fig.~\ref{boxplot}, where the values in the table are the medians of the confidence in the corresponding patch. Denote the intensity and sturctural confidence value in patch $K$ as $C_{int}(K)$ and $C_{str}(K)$ respectively. By design, it should follow $C_{int}(A)>C_{int}(C)>C_{int}(B)$, and $C_{str}(A)\approx C_{str}(C)>>C_{str}(B)$. 
Our structural confidence median values are around 0.6, being much lower than the values in $A$ and $C$, which are close to 1. It indicates that our structural confidence successfully differentiate the artifact patches from non-artifact patches. Also, our intensity confidence have the lowest value in $B$ and highest in $C$, successfully modeling the attenuation. 

\begin{table}[htb]
    \centering
    \begin{tabularx}{\linewidth}{ c *{6}{C} }
    \toprule
    & \multicolumn{3}{c}{reverberation}
            & \multicolumn{3}{c}{shadows}               \\
    & \multicolumn{1}{c}{A} & \multicolumn{1}{c}{C} & \multicolumn{1}{c}{B} & \multicolumn{1}{c}{A} & \multicolumn{1}{c}{C} & \multicolumn{1}{c}{B}             \\
    \midrule
\cite{karamalis2012ultrasound}'s intensity   &   0.90  &   0.17  &   0.15  &   0.30  &   0.33    &   0.61  \\
our intensity   &   0.92  &   0.75  &   0.49  &   0.71  &   0.46    &   0.27  \\
our structural   &   0.96  &   0.98  &   0.65  &   0.96  &   0.98    &   0.61  \\
    \bottomrule
\end{tabularx}
\caption{Quantitative comparison, where in intensity confidence maps, by design, the value should follow $C_{int}(A)>C_{int}(C)>C_{int}(B)$, and in structural confidence maps, $C_{str}(A)\approx C_{str}(C)>>C_{str}(B)$.}
\label{tab}
\end{table}

\subsection{Compounding}
\begin{figure}[h!]

  \centering
  \centerline{\includegraphics[width=8.5cm]{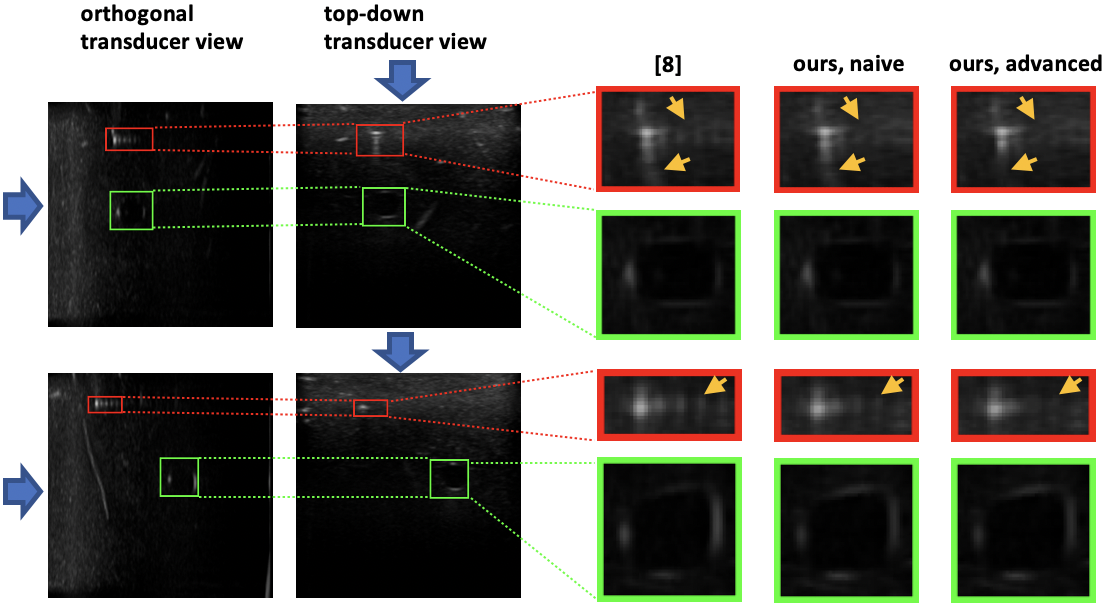}}
\caption{Comparison between using different confidence maps in \cite{zu2014orientation}.  Two inputs are from orthogonal viewpoints, where the blue arrows indicate the probe direction. The zoomed in results on needles and vessels using confidence map by \cite{karamalis2012ultrasound}, our intensity confidence (naive), our intensity confidence while modeling the needle and reverberation (advanced) are shown on the right, where yellow arrows indicate the artifacts.}
\label{comp2}
\end{figure}

We also show the applicability of our results by compounding. Inspired by the uncertainty-based fusion method proposed by \cite{zu2014orientation}, we replace the uncertainty measurement in their method with our confidence map. We compare the compounding results using our confidence maps in uncertainty-based fusion against the original method \cite{zu2014orientation}. The results of compounding two images taken from orthogonal viewpoints are shown in Fig.~\ref{comp2}. Our intensity confidence map performs better in preserving vessel boundaries and removing reverberation artifacts. Explicit modeling of needle and reverberation artifacts allows compounding to better remove reverberation artifacts as shown in column 3 of Fig.~\ref{comp2}, where the reverberation dots are suppressed beside/below the actual needles.

\section{Conclusion}
We developed a new method to model the pixel confidence in ultrasound images. Our pixel-intensity confidence is robust across different tissues and lead to good results in image-compounding algorithms, dealing with the complexities of sound attenuation and diffraction 
Our Structural confidence can be further used to deepen the understanding of ultrasound images, such as shadowing and  reverberation artifacts, which can potentially be used to guide clinicians and surgical robots.

\section{Compliance with Ethical Standards}
\label{sec:ethics}

This study was performed in line with the principles of the Declaration of Helsinki.  The porcine studies were conducted under Pitt IACUC-approved protocol 19014099, as approved by USAMRDC ACURO.

\section{Acknowledgments}
\label{sec:acknowledgments}
This work was sponsored in part by US Army Medical contracts W81XWH-19-C0083, W81XWH-19-C0101, and W81XWH-19-C-0020, and by a PITA grant from the state of Pennsylvania DCED C000072473.  We would like to thank our collaborators at the University of Pittsburgh, Triton Microsystems, Inc., Sonivate Medical, URSUS Medical LLC, and Accipiter Systems, Inc.  We are pursuing intellectual-property protection.  Galeotti serves on the advisory board for Activ Surgical, Inc., and he is a Founder and Director for Elio AI, Inc.”

\bibliographystyle{IEEEbib}
\bibliography{strings,refs}

\end{document}